\documentclass{article}

\usepackage{graphicx}

\title{Compute and Storage Clouds Using Wide Area High Performance Networks}

\author{Robert L. Grossman \and Yunhong Gu  \protect\\
\and Michael Sabala \and Wanzhi Zhang \protect\\
National Center for Data Mining \protect\\
University of Illinois at Chicago}

\date{January 31, 2008}

\begin{document}
\maketitle

\begin{abstract}

  We describe a cloud based infrastructure that we have developed that
  is optimized for wide area, high performance networks and designed
  to support data mining applications.  The infrastructure consists of
  a storage cloud called Sector and a compute cloud called Sphere.  We
  describe two applications that we have built using the cloud and
  some experimental studies.

\end{abstract}

\section{Introduction}

By a cloud, we mean an infrastructure that provides resources and/or
services over the Internet.  A {\em storage cloud} provides storage
services (block or file based services); a {\em data cloud} provides
data management services (record-based, column-based or object-based
services); and a {\em compute cloud} provides computational services.
Often these are layered (compute services over data services over
storage service) to create a stack of cloud services that serves as a
computing platform for developing cloud-based applications.

Examples include Google's Google File System (GFS), BigTable and
MapReduce infrastructure \cite{Dean:04}, \cite{Ghemawat:03}; 
Amazon's S3 storage cloud, SimpleDB data cloud, and EC2 compute cloud
\cite{Amazon:WS07}; and the
open source Hadoop system \cite{Borthakur:2007}, \cite{Hbase:2007}.

Data clouds provide some important advantages for managing and
analyzing data compared to competing technologies.

First, for the majority of applications, databases are the preferred
infrastructure for managing and archiving data sets, but as the size
of the data set begins to grow larger than a few hundred terabytes,
current databases become less competitive with more specialized
solutions, such as the storage services (e.g., \cite{Ghemawat:03},
\cite{Grossman:DMSSP07}) that are parts of data clouds.  For example,
Google's GFS manages Petabytes of data \cite{Dean:07}

Second, data in a data cloud can easily be replicated. Temporary
replicas can be used to improve performance by exploiting locality and
caches, permanent replicas can be used for backing up data, and
long-term replicas can be used for archiving data.  Replicas are
typically placed within a rack, across racks, and across data centers
to handle various types of failures.  Automatic services ensure that
after a failure drops a replica, an additional replica is created.  In
addition, once replicated, the replicated data provides a natural way
to parallelize embarrassingly parallel computations in the cloud.

Third, once data is stored in a cloud, the data can wait for
computing tasks.  In contrast, in a standard grid computing
environment, the data is scattered to nodes in clusters when a
sufficiently large pool of nodes are available; and, in this sense,
the nodes wait for the data. For large data sets, transporting the
data to the nodes can be a significant percentage of the total
processing time.

\begin{figure}
\begin{center}
\begin{tabular}{|c|c|c|} \hline
Application 1 & $\cdots$ & Application $n$ \\ \hline
\multicolumn{3}{|c|}{Cloud-based Compute Services} \\ \hline
\multicolumn{3}{|c|}{Cloud-based Data Services} \\ \hline
\multicolumn{3}{|c|}{Cloud-based Storage Services} \\ \hline
\end{tabular}
\end{center}
\caption{A data stack for a cloud consists of three layered
services as indicated.}
\label{figure:datastack}
\end{figure}

In this paper, we describe a cloud based infrastructure that is
optimized for high performance, wide area networks and designed to
support the ingestion, data management, analysis, and distribution of
large terabyte size data sets.  We assume an ``OptIPuter'' style
design in the sense that we assume that geographically distributed
nodes running storage services are connected by a 10+ Gbps network
that functions more or less as a wide area ``back-plane or bus''.

This paper is organized as follows: Section 2 describes related work.
Section 3 describes a storage cloud called Sector.  Section 4
describes a compute cloud that we have developed called Sphere.  Section 5
describes two Sector and Sphere applications. Section 6 is the summary
and conclusion.

\section{Related Work}

The most common platform for data mining is a single workstation.
There are also several data mining systems that have been developed
for local clusters of workstations, distributed clusters of
workstations and grids \cite{Grossman:SSP03}.  More recently, data
mining systems have been developed that use web services and, more
generally, a service oriented architecture.  For a recent survey of
data mining systems, see \cite{Kargupta:2008}.

By and large, data mining systems that have been developed to date for
clusters, distributed clusters and grids have assumed that the
processors are the scarce resource, and hence shared.  When processors
become available, the data is moved to the processors, the computation
is started, and results are computed and returned \cite{Foster:Grid2}.
To simplify, this is the supercomputing model, and, in the distributed
version, the Teragrid model \cite{Reed:2003}.  In practice with this
approach, for many computations, a good portion of the time is spent
transporting the data.

An alternative approach has become more common during the last few
years.  In this approach, the data is persistently stored and
computations take place over the data when required.  In this model,
the data waits for the task or query.  To simplify, this is the data
center model (and in distributed data version, the distributed data
center model).  The storage clouds provided by Amazon's S3
\cite{Amazon:S3}, the Google File System \cite{Ghemawat:03}, and the
open source Hadoop Distributed File System (HDFS)
\cite{Borthakur:2007} support this model.

To date, work on data clouds \cite{Ghemawat:03, Borthakur:2007,
  Amazon:S3} has assumed relatively small bandwidth between the
distributed clusters containing the data.  In contrast, the Sector
storage cloud described in Section~\ref{section:sector} is designed
for wide area, high performance 10 Gbps networks and employs
specialized protocols such as UDT \cite{Grossman:CN2007} to utilize
the available bandwidth on these networks.

The most common way to compute over GFS and HDFS storage clouds is to
use MapReduce \cite{Dean:04}.  With MapReduce: i) relevant data is
extracted in parallel over multiple nodes using a common ``map''
operation; ii) the data is then transported to other nodes as required
(this is referred to as a shuffle); and, iii) the data is then
processed over multiple nodes using a common ``reduce'' operation to
produce a result set.  In contrast, the Sphere compute cloud described
in Section~\ref{section:sphere} allows arbitrary user defined
operations to replace both the map and reduce operations.  In
addition, Sphere uses specialized network transport protocols
\cite{Grossman:CN2007} so that data can be transferred efficiently
over wide area high performance networks during the shuffle operation.

\section{Sector Storage Cloud}
\label{section:sector}

\begin{figure}
\begin{center}
\begin{tabular}{|c|} \hline
{\bf File Locating and Access Services} \\ \hline
{\bf Distributed Storage Services} \\ \hline
{\bf Routing Services} \\ \hline
\end{tabular}
\end{center}
\caption{A Sector Server provides file locating and file access
  services to any Sector Client.  Sector maintains multiple copies of
  files, locates them using a P2P-based routing protocol, and transfers
  them using specialized network protocols such as UDT.  Sector is
  layered so that other routing protocols may be used.}
\label{figure:sector-server}
\end{figure}

Sector has a layered architecture: there is a routing layer and a
storage layer.  Sector services, such as the Sphere compute cloud
described below, are implemented over the storage layer.  See
Figure~\ref{figure:sector-server}.

The routing layer provide services that locate the node that stores
the metadata for a specific data file or computing service.  That is,
given a name, the routing layer returns the location of the node that
has the metadata, such as the physcial location in the system, of the
named entity.  Any routing protocols that can provide this function
can be deployed in Sector. Currently, Sector uses the Chord P2P
routing protocol \cite{Stoica:2001}.  The next version of Sector will
support specialized routing protocols designed for uniform wide area
clouds, as well as non-uniform clouds in which bandwidth may vary
between portions of the cloud.

Data transport itself is done using specialized high performane
network transport protocols, such as UDT \cite{Grossman:CN2007}.
UDT is a rate-based  application layer network transport protocol 
that supports large data flows over wide area high performance networks.
UDT is {\em fair} to several large data flows in the sense that it shares
bandwidth equally between them.  UDT is also {\em friendly} to TCP flows
in the sense that it backs off, enabling any TCP flows sharing the network
to use the bandwidth they require.

The storage layer manages the data files and their metadata. It
maintains an index of the metadata of files and creates replicas.
A typical Sector data access session involves the following steps:
\begin{enumerate}

\item The Sector client connects to any known Sector server S, and
  requests the locations of a named entity.

\item S runs a look-up inside the server network using the services
  from the routing layer and returns one or more locations to the
  client.

\item The client requests a data connection to one or more servers on
  the returned locations using a specialized Sector library designed
  to provide efficient message passing between geographically
  distributed nodes.

\item All further requests and responses are performed using UDT over
  the data connection established by the message passing library.

\end{enumerate}

\begin{figure}
\begin{center}
\includegraphics[scale=0.4]{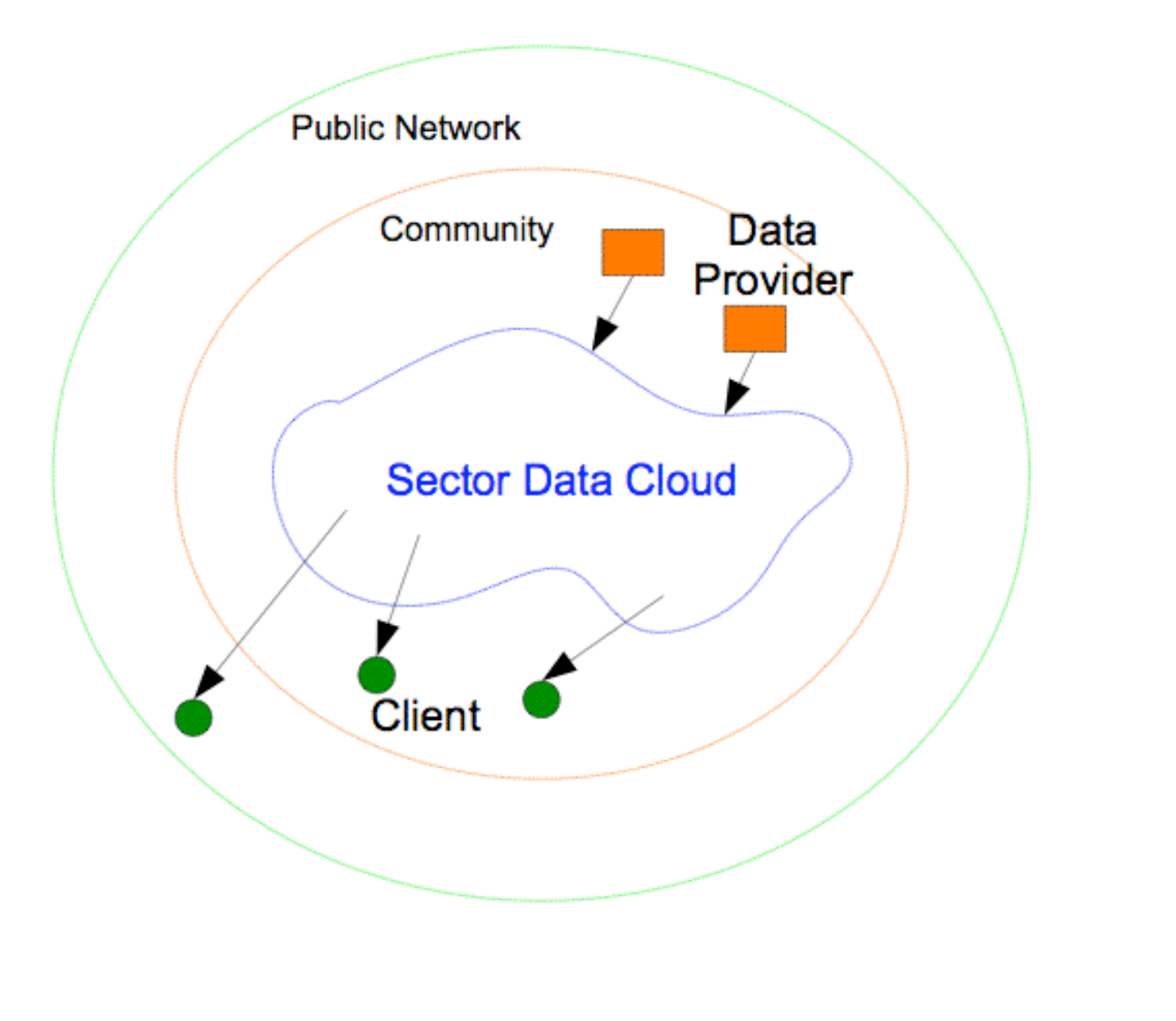}
\end{center}
\caption{With Sector, only users in a community who have been added to
  the Sector access control list can write data into Sector.  On the
  other hand, any member of the community or of the public can read
  data, unless additional restrictions are imposed.}
\label{figure:sector-access}
\end{figure}

Figure~\ref{figure:sector-access} depicts a simple Sector
system. Sector is meant to serve a community of users. The Sector
server network consists of nodes managed by administrators within the
community.  Anyone within the community who has an account can write
data to the system.  In general, anyone in the public can read data
from Sector.  In contrast, systems such as GFS \cite{Ghemawat:03} and
Hadoop \cite{Borthakur:2007} are targeted towards organizations (only
users with accounts can read and write data), while systems such as
Globus \cite{Foster:Grid2} are targeted towards virtual organizations
(anyone with access to a node running GSI \cite{Foster:Grid2} and
having an account can read and write data).  Also, unlike some
peer-to-peer systems, while reading data is open, writing data in
Sector is controlled through access control lists.

Below are the typical Sector operations:
\begin{itemize}


\item The Sector storage cloud is automatically updated
when nodes join or leave the cloud --- it is not required
that this be done through a centralized control system.

\item Data providers within the community who have been added to the
  access control lists can upload data files to Sector Servers.

\item The Sector Storage Cloud automatically creates data replicas for
long term archival storage, to provide more efficient content distribution,
and to support parallel computation.

\item Sector Storage Cloud clients can connect to any known server
  node to access data stored in Sector. Sector data is open to the
  public (unless further restrictions are imposed). 


\end{itemize}

Using P2P for distributing large scientific data has becomed popular recently. Some related work can be found in \cite{FR:FGCS2008}, \cite{DAM:FGCS2008}, and \cite{Tru:FGCS2007}.

\section{Sphere}
\label{section:sphere}

Sphere is middleware that provides distributed computing services for
persistent distributed data managed by Sector.  Sphere is designed to
perform computations over data without moving it whenever possible.
If an application uses the Sphere client API, then Sphere provides the
following services: locating data, moving data (if required), locating
and managing computing resources, load balancing, and fault tolerance.
The distributed parallelization is done implicitly by Sphere: Sphere
automatically locates computing nodes to run the processing function
in parallel, while Sector provides a uniform data access interface.

Sphere is broadly based upon the stream approach to data processing in
the sense that all processing assumes that each record in the stream
is processed independently by the same processing function.  More
specifically, the data is viewed as a stream that is divided into
chunks.  The chunks are already naturally distributed over the nodes
managed by the Sector Storage Cloud.  A Sphere application provides
one or more processing functions that are applied to each record in
the data stream independently.  Sphere automatically invokes the
processing function over multiple nodes in parallel.  After a
processing stage, data can be transfered from node to node as
required.  The cycle then repeats by applying another data processing
function, followed by another data transfer.

As an example, consider the following loop in a serial program.

\begin{verbatim}
   for (int i = 0; i < 100000000; ++ i)
      process(data[i]);
\end{verbatim}

In the stream process paradigm used by Sphere, this loop will be replaced by: 

\begin{verbatim}
   sphere.run(data, process);
\end{verbatim}

The majority of the processing time for many data intensive
applications is spent in loops like these; developers often spend a
lot of time parallelizing these types of loops using MPI or PVM.
Parallelizing these loops in distributed environments presents
additional challenges. Sphere provides a simple way for application
developers to express these loops and then automatically parallelizes
and distributes the required computations.

The approach taken by Sphere is to provide a very simple distributed
application development interface by limiting the type of operations
supported. The stream processing paradigm used in Sphere is
fundamentally a simplified data parallel and master/worker pattern.

Although the stream processing paradigm is a special-purpose parallel
computing model, it has been used successfully in general purpose GPU
programming (GPGPU). Google's MapReduce system \cite{Dean:04} also
uses the stream processing paradigm to process very large data sets
managed by the Google File System (GFS) \cite{Ghemawat:03}.

\section{Sector/Sphere Applications}
\label{section:applications}

\subsection{Experimental Setup}

The applications described in this section run on a wide area, high
performance testbed called the Teraflow Testbed
\cite{Grossman:TFT2007}.  The various sites on the testbed are
connected using 10 Gbps networks.  Each site contains a small cluster
of 2 -- 16 dual dual-core (i.e., total 4-core) Opteron servers.  There
are sites in Chicago, Pasadena (CA, USA), McLean (VA, USA), Greenbelt
(MD, USA), Tokyo (Japan), Daejeon (Korea). Each Opteron server has a
2.4Ghz CPU and 4GB memory.  The furthest two nodes in the testbed have
a RTT of 200ms between them.

\subsection{Distributing the Sloan Digital Sky Survey (SDSS)}

One of the the first applications we developed over Sector was a
content distribution network for large e-science data sets.  In
particular, we have used the Sector Cloud to distribute the Sloan
Digital Sky Survey data \cite{Szalay:Science01} to astronomers around
the world.

The SDSS data consists of the 13TB DR5 data release (60 catalog files,
64 catalog files in EFG format, 257 raw image data collection files)
and the 14TB DR6 data release (60 catalog files, 60 Segue files, 268
raw image collection files). The sizes of these files range between
5GB and 100GB each.

The Sector Cloud has been used to distributed the SDSS since July
2006. During the last 18 months, we have had about 5000 system
accesses and a total of 200TB of data was transferred to end users.

In order to evaluate Sector's wide area data transfer performance we
defined a measure called LLPR, or {\em long distance to local
  performance ratio}, which is the ratio of the performance measured
over the wide area network divided by the performance over a local
area network containing machines with the same configuration.

The higher the LLPR, the better the performance.  The maximum possible
performance is when the LLPR is equal to 1.0. That is, a long distance
data transfer cannot be faster than a local transfer with the same
hardware configuration.

\begin{table}
\begin{tabular}{|l|l|l|l|}\hline
{\bf Source} & {\bf Destination} & {\bf Throughput (Mb/s)} & {\bf LLPR} \\ \hline
Greenbelt,MD &	Daejeon, Korea  & 360  & 0.78 \\ \hline
Chicago, IL  & Pasadena, CA  & 550 & 0.83 \\ \hline
Chicago, IL  & Greenbelt, MD  & 615 & 0.98 \\ \hline
Chicago, IL  & Tokyo, Japan  & 490 & 0.61 \\ \hline
Tokyo, Japan & Pasadena, CA  & 550 & 0.83 \\ \hline
Tokyo, Japan  & Chicago, IL  & 460 & 0.67 \\ \hline
\end{tabular}
\caption{This table shows that the Sector Storage Cloud provides access to large terabyte
size e-science data sets at 0.60\% to 0.98\% of the performance that would
be available to scientists sitting next to the data.}
\end{table}

\subsection{Identifying Emergent Behavior in TCP/IP Traffic}

Angle is a Sphere application that identifies anomalous or suspicious
behavior in TCP packet data that is collected from multiple,
geographically distributed sites \cite{Grossman:NGDM2007}.  Angle
contains Sensor Nodes that are attached to the commodity Internet and
collect IP data. Connected to each Sensor Node on the commodity
network is a Sector node on a wide area high performance network.  The
Sensor Nodes zero out the content, hash the source and destination IP
to preserve privacy, package moving windows of anonymized packets in
pcap files \cite{Beale:2007} for further processing, and transfer
these files to its associated Sector node. Sector services are used to
manage the data collected by Angle and Sphere services are used to
identify anomalous or suspicious behavior.

Angle Sensors are currently installed at four locations: the
University of Illinois at Chicago, the University of Chicago, Argonne
National Laboratory and the ISI/University of Southern California.
Each day, Angle processes approximately 575 pcap files totaling
approximately 7.6GB and 97 million packets.  To date, we have
collected approximately 140,000 pcap files.

For each pcap file, we aggregate all the packet data by
source IP (or other specified entity), compute features, and then
cluster the resulting points in feature space.  With this process
a model summarizing a cluster model is produced for each pcap file.

Through a temporal analysis of these cluster models, we identify 
anomalous or suspicious behavior and send appropriate alerts.
See \cite{Grossman:NGDM2007} for more details.

Table~\ref{fig:shere-exp} shows the performance of Sector and Sphere
when computing cluster models using the k-means algorithms
\cite{Kumar:DMBook2006} from distributed pcap files ranging in size
from 500 points to 100,000,000.

\begin{table}
\begin{center}
\begin{tabular}{|p{1.275in}|p{1.275in}|p{0.75in}|} \hline 
{\bf Number records} & {\bf Number of Sector Files} & {\bf Time} \\ \hline
500 & 1 & 1.9 s \\ \hline
1000 & 3 & 4.2 s \\ \hline
1,000,000 & 2850 & 85 min \\ \hline
100,000,000 & 300,000 & 178 hours \\ \hline
\end{tabular}
\end{center}
\caption{The time spent clustering using Sphere scales as the number of
records increases, as is illustrated in the table above from 500 records to 100,000,000
records.}
\label{fig:shere-exp}
\end{table}

\subsection{Hadoop vs Sphere}
In this section we describe some comparisons between Sphere and Hadoop
\cite{Borthakur:2007} on a 6-node Linux cluster in a single location.
We ran the TeraSort benchmark \cite{Borthakur:2007} using both Hadoop
and Sphere.  The benchmark creates a 10GB file on each node and then
performs a distributed sort.  Each file contains 100-byte records with
10-byte random keys.

The file generation required 212 second per file per node for Hadoop,
which is a throughput of 440Mbps per node. For Sphere, the file
generation required 68 seconds per node, which is a throuhgput of
1.1Gbps per node.

Table~\ref{fig:hs-sort} shows the performance of the sort phase (time
in seconds). In this benchmark, Sphere is significantly faster
(approximately 2 to 3 times) than Hadoop.  It is also important to
note that in this experiment, Hadoop uses all four cores on each node,
while Sphere only uses one core.

\begin{table}
\begin{center}
\begin{tabular}{|p{0.50in}|p{0.50in}|p{0.50in}|p{0.50in}|p{0.50in}|p{0.50in}|p{0.50in}|}\hline
{\bf Node} & {\bf 1} & {\bf 2}  & {\bf 3} & {\bf 4} & {\bf 5} & {\bf 6}\\ \hline 
Hadoop & 1708 & 1801 & 1850 & 1881 & 1892 & 1953 \\ \hline
Sphere & 510 & 820 & 832 & 850 & 866 & 871 \\ \hline 
\end{tabular}
\end{center}
\caption{This table compares the performance of Sphere and Hadoop
sorting a 10GB file on each of six nodes.  The time is in seconds.} 
\label{fig:hs-sort}
\end{table}

\section{Summary and Conclusion}

Until recently, most high performance computing relied on a model in
which cycles were scarce resources that were managed and data was
moved to them when required.  As data sets grow large, the time
required to move data begins to dominate the computation.  In
contrast, with a cloud-based architecture, a storage cloud provides
long-term archival storage for large data sets.  A compute cloud is
layered over the storage to provide computing cycles when required and
the distribution and replication used by the storage cloud provide a
natural framework for parallelism.

In this paper, we have described a high performance storage cloud
called Sector and a compute cloud called Sphere that are designed to
store large distributed data sets and to support the parallel analysis
of these data sets.  Sector and Sphere rely on specialized high
performance data transport protocols such as UDT that use bandwidth
efficiently over wide area high bandwidth networks.  

We have also described two applications that use this infrastructure
and shown that with wide area high performance networks and a
cloud-based architecture that computing with distributed data can be
done with approximately the same efficiency as computing with local
data.

\section*{Acknowledgments}

This work was supported in part by the National Science Foundation
through grants SCI-0430781, CNS-0420847, and ACI-0325013.

\end{document}